\documentclass[prb,twocolumn,showpacs]{revtex4}

\usepackage[dvipdfmx]{graphicx}
\usepackage{amsmath,amssymb,amsfonts}
\usepackage{bm}
\begin{document}

\title{Phonon, Two-Magnon and Electronic Raman Scattering of Fe$_{1+y}$Te$_{1-x}$Se$_x$} 

\author{K.~Okazaki$^{1,4}$}
\altaffiliation{Present address: Institute for Solid State Physics, University of Tokyo, Kashiwa, Chiba 277-8581, Japan}
\email{okazaki@issp.u-tokyo.ac.jp}
\author{S.~Sugai$^{1,4}$}
\author{S.~Niitaka$^{2,4}$}
\author{H.~Takagi$^{2,3,4}$} 
\affiliation{
$^1$Department of Physics, Nagoya University, Nagoya, Aichi 464-8602, Japan\\
$^2$RIKEN (The Institute of Physical and Chemical Research), Wako, Saitama 351-0198, Japan\\
$^3$Department of Advanced Materials Science, University of Tokyo, Kashiwa, Chiba 277-8581, Japan\\
$^4$JST, Transformative Research-Project on Iron Pnictides (TRIP), Chiyoda-Ku, Tokyo 102-0075, Japan
}

\date{\today} 

\begin{abstract}
We have measured Raman scattering spectra of single-crystalline FeTe$_{0.6}$Se$_{0.4}$ ($T_c \sim$ 14.5 K) and its parent compound Fe$_{1.074}$Te at various temperatures. In the parent compound Fe$_{1.074}$Te, $A_{1g}$ and $B_{1g}$ modes have been observed at 157.5 and 202.3 cm$^{-1}$, respectively, at 5 K. These frequencies qualitatively agree with the calculated results. Two-magnon excitation has been observed around 2300 cm$^{-1}$ for both compounds. Temperature dependence between the electronic Raman spectra below and above $T_c$ has been observed and 2$\Delta$ and 2$\Delta$/$k_BT_C$ have been estimated as 5.0 meV and 4.0, respectively. 
\end{abstract}


\maketitle

\section{Introduction}
Since the discovery of superconductivity in LaFeAsO$_{1-y}$F$_y$  (1111 system) with $T_c$ = 26 K,~\cite{Kamihara2008JACS} extensive researches has been devoted to the iron-based superconductors. Immediately after the discovery, $T_c$ enhancement of LaFeAsO$_{1-y}$F$_y$ under high-pressure upto $\sim$ 43 K~\cite{Takahashi2008Nature} and higher $T_c$ ($\sim$ 55 K) of SmFeAsO$_{1-y}$F$_y$~\cite{Ren2008CPL} has been found. Up to now, various iron-based superconductors with the different crystal structures has been discovered such as Ba$_{1-x}$K$_x$Fe$_2$As$_2$ (122 system),~\cite{Rotter2008PRB,Rotter2008PRL} LiFeAs (111 system),~\cite{Tapp2008PRB,Wang2008SSC} Fe$_{1+y}$Te$_{1-x}$Se$_x$ (11 system),~\cite{Hsu2008PNAS,Yeh2008EPL,Fang2008PRB} Sr$_4$V$_2$O$_6$Fe$_2$As$_2$ (42622 system),~\cite{Zhu2009PRB} and so on. These iron-based superconductors commonly have Fe$Pn_4$ ($Pn$: pnictogen) or Fe$Ch_4$ ($Ch$: chalcogen) tetragonal layers. In addition, there are some other common features in the iron-based superconductors, i.e., antiferromagnetic (AFM) ordering in the parent compound, nesting feature of the Fermi surfaces (FS), and so on. FS of these iron-based superconductors have been calculated on the basis of the density functional theory (DFT)~\cite{Singh2008PRL,Singh2008PRB,Subedi2008PRB,Wang2009PRB} and it has been pointed that hole FS around $\Gamma$ point and electron FS around M point are nearly nested. This has been confirmed by the calculations of Lindhard response function.~\cite{Singh2008PRB,Dong2008EPL,Xu2008EPL,Deng2009PRB,Wang2009PRB} The importance of spin fluctuations due to the nesting feature of FS and a possibility of s$_\pm$ superconducting-gap symmetry have been proposed.~\cite{Mazin2008PRL,Kuroki2008PRL} Experimentally, FS similar to the DFT results~\cite{Ding2008EPL,Xia2009PRL} and AFM ordering in the parent compounds~\cite{Cruz2008Nature,Huang2008PRL,Li2009PRB} have been observed. 

Superconductivity of 11 system has first discovered by Hsu {\it et al.}~\cite{Hsu2008PNAS} in FeSe with $T_c \sim$ 8 K. $T_c$ of substituted compounds reaches up to $\sim$ 14 K~ at $x$ = 0.4.~\cite{Fang2008PRB} Under hydrostatic pressure, $T_c$ of Fe$_{1+y}$Se is enhanced to $\sim$ 37 K around 7-9 GPa.~\cite{Medvedev2009NM,Margadonna2009PRB} Among the iron-based superconductors, 11 system has the simplest structure and consists of only Fe$Ch_4$ layer.~\cite{Margadonna2008CC} Hence, 11 system can be assumed to be suitable to reveal the relationship between the crystal structure and the origin of superconductivity. However, important differences between 11 system and the other iron-arsenide superconductors should be noted, i.e., while the iron arsenide superconductors show single-stripe AFM ordering~\cite{Cruz2008Nature,Huang2008PRL,Li2009PRB} and those AFM wave vector $Q_{AF}$ = ($\pi$,$\pi$) coincide with the $\Gamma$-$M$ nesting vector between the hole FS around $\Gamma$ point and the electron FS around $M$ point. However, Fe$_{1+y}$Te shows a double-stripe AFM ordering and its $Q_{AF}$ = ($\delta\pi$, 0) is rotated by 45$^\circ$ from the nesting vector.~\cite{Li2009PRB,Bao2009PRL} $\delta$ is tunable depending on the amount of the excess irons and can be incommensurate for the large $y$.~\cite{Bao2009PRL} Hence, 11 system can be regarded as a counter example that the nesting condition is not crucial for the AFM ordering. Hence, we may have to reconsider the relationship between the superconductivity and the spin fluctuation. On the other hand, an inelastic neutron scattering (INS) measurement in FeTe$_{0.6}$Se$_{0.4}$ reveals a resonance feature below $T_c$ with the excitation energy of 6.5 meV and the wave vector of ($\pi$,$\pi$).~\cite{Qiu2009PRL} Since this wave vector corresponds to the nesting vector, there is a possibility that while the long range AFM ordering in the parent compound of 11 system is not related to the FS nesting, the spin fluctuation with the ($\pi$,$\pi$) nesting wave vector is crucial for the emergence of superconductivity.

In this paper, we have reported the phonon, two-magnon, and electronic Raman scattering of the 11 system Fe$_{1.074}$Te and FeTe$_{0.6}$Se$_{0.4}$ at various temperatures. Because the parent compound Fe$_{1.074}$Te has a simultaneous structural and magnetic transition, some anomaly around the transition temperature can be observed in the phonon and/or magnon Raman scattering. On the other hand, it is interesting that how the Se-substitution effects are observed in the two-magnon excitation of the parent AFM compound Fe$_{1.074}$Te. This could be an important clue to reveal the relationship between the magnetic fluctuations and superconductivity. From the electronic Raman spectra, the size and symmetry of the superconducting gap could be determined. 

\section{Experimental}
\begin{figure} [t]
\begin{center}
\includegraphics[width=9cm]{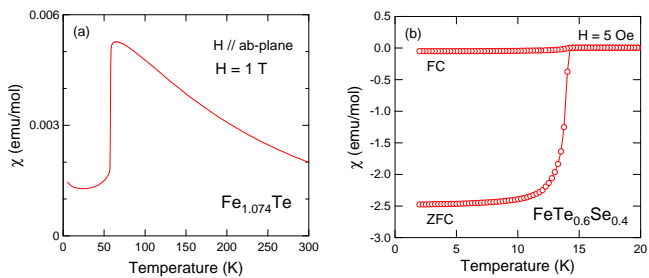} 
\caption{(Color online) Results of magnetic measurements for (a) Fe$_{1.074}$Te and (b) FeSe$_{0.6}$Te$_{0.4}$. $T_s$ and $T_c$ have been confirmed as $\sim$ 58 K and 14.5 K, respectively.}
\label{Fig1} 
\end{center}
\end{figure}

Single crystals of Fe$_{1+y}$Te$_{1-x}$Se$_{x}$ were grown by a melt-growth technique. Nominal compositions of the grown crystals were FeTe$_{0.9}$ and FeTe$_{0.5}$Se$_{0.5}$. The detailed procedures have been described in Ref.~\onlinecite{Hanaguri2010Science}. The actual compositions were confirmed  by the inductively coupled plasma (ICP) analysis as Fe$_{1.074}$Te and FeTe$_{0.6}$Se$_{0.4}$, respectively. The structural and magnetic transition temperature $T_s$ of the parent compound Fe$_{1.074}$Te and the superconducting transition temperature $T_c$ of FeSe$_{0.6}$Te$_{0.4}$ were confirmed by a
superconducting quantum interference device (SQUID) magnetometer as shown in Fig.~\ref{Fig1}. $T_s$ and $T_c$ were $\sim$ 58 K and 14.5 K, respectively. Raman-scattering spectra were measured with the fresh cleaved surfaces in a quasibackscattering configuration using a 5145 {\AA} Ar ion laser, a triple monochromator, and a liquid-nitrogen-cooled CCD detector. The laser power was set to 20 mW and the laser spot was focused to 50$\times$500 $\mu$m$^2$ on the sample surface. The wide-energy spectra were obtained by shifting the central wave number of the spectrometer. The obtained spectra were corrected for the efficiency of the spectrometer utilizing a standard lamp to keep the constant response for the light power. The polarization configuration is denoted by $k_i(E_i,E_s)k_s$, where $k_i$($k_s$) is the direction of the wave vector of incident (scattered) light and $E_i$($E_s$) is the polarization of the incident(scattered) light. The spectra were measured at four polarization configurations $c(aa)\bar{c}$, $c(ab)\bar{c}$, $c(xx)\bar{c}$, and $c(xy)\bar{c}$ in the $ab$-plane, where $a$ and $b$ denote the crystallographic $a$- and $b$-axis, respectively and $x$ and $y$ are the directions rotated by 45$^\circ$ from $a$-axis and $b$-axis in the $ab$-plane, respectively. The Raman active symmetries are $A_{1g}+B_{1g}$, $B_{2g}$, $A_{1g}+B_{2g}$, and $B_{1g}$ for $(aa)$, $(ab)$, $(xx)$, and $(xy)$ polarization configurations, respectively. 

\section{Calculation}
We have also performed calculations of the phonon modes of FeSe and FeTe in a non-magnetic tetragonal phase in the framework of the density-functional perturbation theory based on the plane-wave basis method using the Quantum-Espresso code.~\cite{QE} The lattice parameters $a$ = 3.8097 {\AA} and $c$ = 6.2756 {\AA} for FeTe and $a$ = 3.7696 {\AA} and $c$ = 5.520 {\AA} for FeSe were used according to the X-ray diffraction measurements by Mizuguchi {\it et al.}~\cite{Mizuguchi2009JPSJ} The internal parameter $z$ corresponding to the chalcogen height from the iron plane was optimized via energy minimization. The optimized parameter $z$ was 0.2507 for FeTe and 0.2343 for FeSe, respectively. These values are almost the same with the those used by Subedi {\it et al.}~\cite{Subedi2008PRB} The cutoff energies were 50 Ry for the wavefunctions and 500 Ry for the electron density, respectively. 

\section{Results and Discussion}

\subsection{Phonon Raman Scattering}

\begin{figure} [t]
\begin{center}
\includegraphics[width=8.5cm]{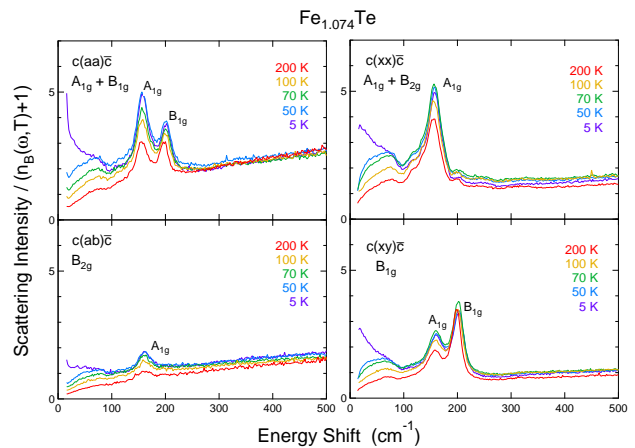} 
\caption{(Color online) Temperature-dependent phonon Raman spectra of Fe$_{1.074}$Te. $A_{1g}$ and $B_{1g}$ modes have been observed at 157.5 cm$^{-1}$ and 202.3 cm$^{-1}$, respectively, at 5 K.}
\label{Fig2} 
\end{center}
\end{figure}

From the group theory considerations, $\Gamma$-point phonon modes of the tetragonal Fe(Te,Se) can be expressed as $\Gamma = A_{1g} + 2A_{2u} + B_{1g} + 2E_{g} + 2E_{u}$. While the Raman-active modes $\Gamma_{Raman} = A_{1g} + B_{1g} + 2E_{g}$, $A_{1g}$ and $B_{1g}$ modes can be observed in our measurements. Figure~\ref{Fig2} shows phonon Raman spectra of Fe$_{1.074}$Te in all the measured configurations. From the polarization dependence, we have concluded that $A_{1g}$ and $B_{1g}$ modes have been observed at 158 cm$^{-1}$ and 202 cm$^{-1}$, respectively, at 5 K. Although $A_{1g}$ mode should not be observed in the $(ab)$ and $(xy)$ configurations, this mode has been observed weakly also in those configurations. This should be due to the leakage of other polarization components. While the peak positions have slightly shifted to the higher energies with decreasing temperature, any distinct change has not been observed across the magnetic and structural transition temperature $T_s$. In other phonon Raman scattering measurements, Xia {\it et al.} have reported that $A_{1g}$ and $B_{1g}$ modes are 159.1 cm$^{-1}$ and 196.3 cm$^{-1}$, respectively, for single-crystal FeTe$_{0.92}$~\cite{Xia2009PRB} and Kumar {\it et al.} have reported that those are 160 cm$^{-1}$ and 224 cm$^{-1}$, respectively, for polycrystalline FeSe$_{0.82}$.~\cite{Kumar2010SSC} 

\begin{table}[t]
\begin{center}
\caption{(Color online) The results of phonon-mode calculation and comparison with the Raman measurements}
\label{Table1}
\begin{tabular}{cccccc} \hline\hline
\multicolumn{1}{c}{Symmetry} & FeTe  & FeSe  & Expt.  & \multicolumn{1}{c}{Atoms} & \multicolumn{1}{c}{Active} \\
\multicolumn{1}{c}{} & (cm$^{-1}$) & (cm$^{-1}$) & Fe$_{1.074}$Te & \multicolumn{1}{c}{} & \multicolumn{1}{c}{} \\
\hline
$E_g$ & 120.1 & 167.1 &  & (Te,Se) & Raman \\
$A_{1g}$ & 168.4 & 221.9 & 158 & (Te,Se) & Raman \\
$B_{1g}$ & 216.4 & 179.8 & 202 & Fe & Raman \\
$E_u$ & 227.4 & 247.5 &  & Fe+(Te,Se) & IR \\
$E_g$ & 274.1 & 313.0 &  & Fe & Raman \\
$A_{2u}$ & 302.8 & 290.9 &  & Fe+(Te,Se) & IR \\\hline\hline
\end{tabular}
\end{center}
\end{table}

\begin{figure} [t]
\begin{center}
\includegraphics[width=8cm]{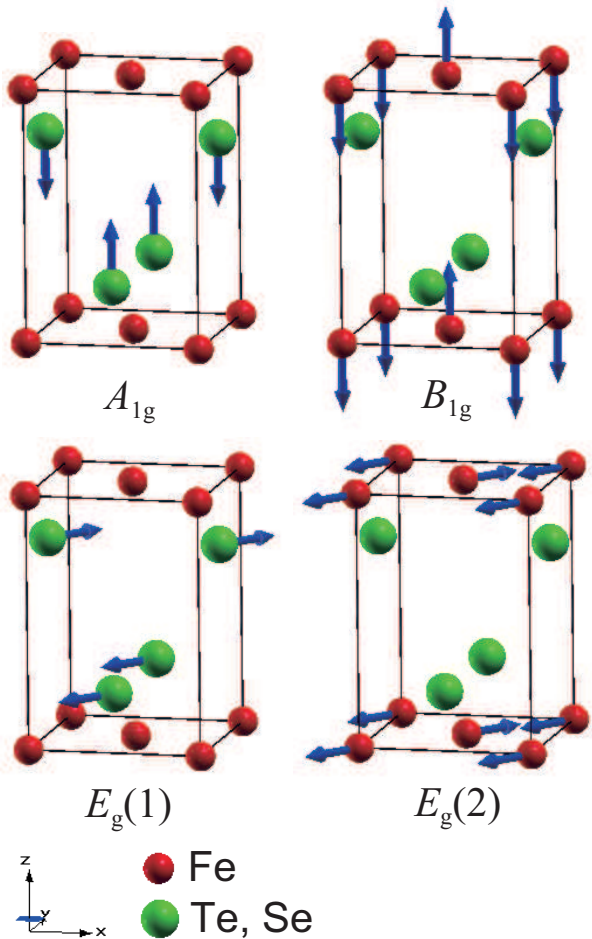} 
\caption{(Color online) Atomic-displacement patterns of Raman active optical modes of Fe(Te,Se).}
\label{Fig3} 
\end{center}
\end{figure}

Table~\ref{Table1} shows a summary of the calculated results for the optical phonons of FeTe and FeSe. The irreducible representations, frequencies, main contribution of displaced atoms, and activities have been listed. The experimental results of Fe$_{1.074}$Te also have been listed. The calculated frequencies of $A_{1g}$ and $B_{1g}$ modes of FeTe are qualitatively in accord with the the observed peak positions. The atomic displacement patterns of Raman-active modes are shown in Fig.~\ref{Fig3}. The observed $A_{1g}$ and $B_{1g}$ modes can be assigned to the $c$-axis anti-phase vibration modes of the chalcogens and the irons, respectively. 

\begin{figure} [t]
\begin{center}
\includegraphics[width=8.5cm]{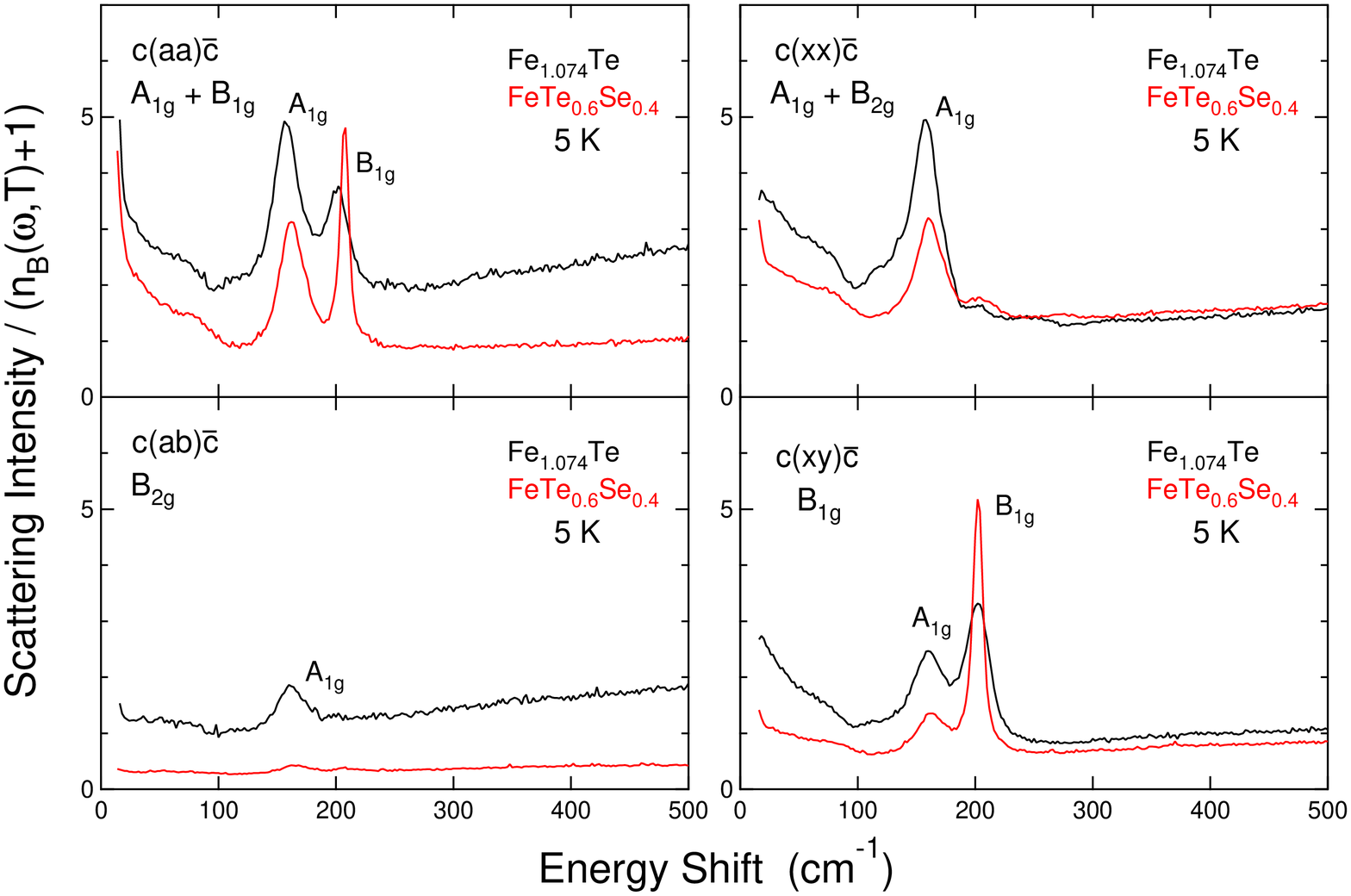} 
\caption{(Color online) Comparison of phonon Raman spectra between the parent compound Fe$_{1.074}$Te and the superconducting FeTe$_{0.6}$Se$_{0.4}$. $A_{1g}$ and $B_{1g}$ modes of FeTe$_{0.6}$Se$_{0.4}$ have been observed at 161 cm$^{-1}$ and 202 cm$^{-1}$, respectively, at 5 K.}
\label{Fig4} 
\end{center}
\end{figure}

Figure~\ref{Fig4} shows a comparison of phonon Raman spectra between the parent compound Fe$_{1.074}$Te and the superconducting FeTe$_{0.6}$Se$_{0.4}$. $A_{1g}$ and $B_{1g}$ modes of FeTe$_{0.6}$Se$_{0.4}$ have been observed at 161 cm$^{-1}$ and 202 cm$^{-1}$, respectively, at 5 K. Because $B_{1g}$ mode is an iron vibration mode, this energy is almost the same between these two compounds. On the other hand, $A_{1g}$ is a chalcogen vibration mode, this energy is slightly larger for FeTe$_{0.6}$Se$_{0.4}$. However, the energy difference of $B_{1g}$ mode between Fe$_{1.074}$Te and FeTe$_{0.6}$Se$_{0.4}$ seems to be too small as compared to the calculated results for FeTe and FeSe. This would be related to the facts that difference of the lattice constants between FeSe and substituted compound FeTe$_{1-x}$Se$_{x}$ are relatively large and that a miscible region exists around $x = 0.7-0.95$~\cite{Mizuguchi2009JPSJ}. Furthermore, the width of $B_{1g}$ mode seems to be much broader for Fe$_{1.074}$Te. This may be due to the inhomogeneity at the iron sites. As far as characterized by ICP, while the parent compound Fe$_{1.074}$Te include excess irons, they cannot be detected in the superconducting FeTe$_{0.6}$Se$_{0.4}$.   

\subsection{Two-magnon Raman Scattering}

\begin{figure} [t]
\begin{center}
\includegraphics[width=8.5cm]{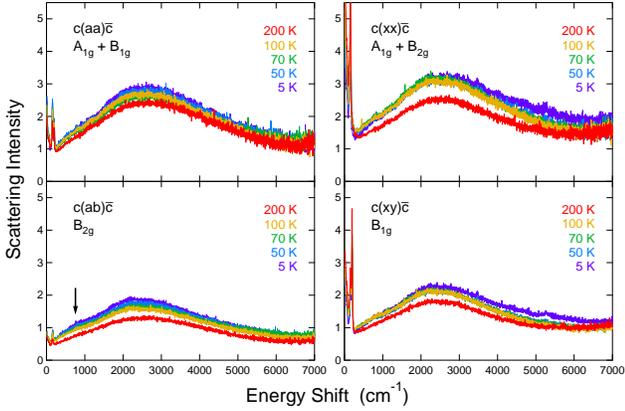} 
\end{center}
\caption{(Color online) Temperature-dependent Raman spectra of Fe$_{1.074}$Te with the wider-energy region up to 7000 cm$^{-1}$. Broad peak structures around 2300 cm$^{-1}$ is assigned to the two-magnon excitation. Broad weak edge structures around 300-800 cm$^{-1}$ observed below $T_s$ are denoted by the arrow in the (ab) configuration spectra and are considered to be related to the magnetic transition.}
\label{Fig5} 
\end{figure}

Figure~\ref{Fig5} shows Raman spectra of Fe$_{1.074}$Te in the wider-energy region up to 7000 cm$^{-1}$ ($\sim$ 870 meV). Broad peak structures have been observed in all the configurations at $\sim$ 2300 cm$^{-1}$ ($\sim$ 285 meV). Furthermore, broad weak edge structures seem to evolve around 300-800 cm$^{-1}$ (37-99 meV) below $T_s$ as denoted by the arrow in the (ab) configuration spectra of Fig.~\ref{Fig5}. We have assigned the broad peak structures around 2300 cm$^{-1}$ to the two-magnon scattering and the weak edge structures to the excitation related to the magnetic transition. In the Raman spectra of BaFe$_2$As$_2$ (Ba122), similar broad peak and weak edge structures have been observed.~\cite{Sugai2010PRB,Sugai2010arXiv} The weak edge structures evolve below the spin-density-wave (SDW) transition temperature $T_{SDW}$ around 400 and 800 cm$^{-1}$ for Ba122. In the optical conductivity $\sigma(\omega)$ of Ba122, also, double-peak structures develop below $T_{SDW}$ and this feature is considered to be related to the SDW gap.~\cite{Hu2008PRL} On the other hand, SDW gap has not been observed in $\sigma(\omega)$ of Fe$_{1.05}$Te.~\cite{Chen2009PRB} However, $\sigma(\omega)$ of Fe$_{1.05}$Te shows a prominent temperature dependence below 500 cm$^{-1}$ below $T_S$. The edge structure around 300-800 cm$^{-1}$ in the Raman spectra of Fe$_{1.074}$Te also should be related to the magnetic transition. 

The energy of two-magnon scattering is also similar between Fe$_{1.074}$Te and Ba122.~\cite{Sugai2010PRB,Sugai2010arXiv} This may be surprising at first glance, because the magnetic structure is different between Fe$_{1.074}$Te and Ba122. While Ba122 has a single-stripe AFM structure with the AFM wavevector {\bf Q}$_{AF}$ = ($\pi$, $\pi$), Fe$_{1+y}$Te has a double-stripe AFM structure with {\bf Q}$_{AF}$ = ($\delta\pi$, 0). It is natural that one can expect different magnetic interactions from the different AFM structures. In the single-stripe AFM structure, there are two kinds of the first-nearest-neighbor exchange interactions, the AFM interaction $J_{1a}$ and the ferromagnetic (FM) interaction $J_{1b}$. The second-nearest-neighbor interaction is only one kind of $J_2$. On the other hand, in the double-stripe AFM structure, the second-nearest-neighbor interaction is also two kinds, the AFM interaction $J_{2a}$ and the FM interaction $J_{2b}$. These notations are according to Han {\it et al.}~\cite{Han2009PRL,Han2009PRLa} They have estimated exchange interactions of various iron-based superconductors. Experimentally, Zhao {\it et al.} have confirmed that the maximum spin-wave bandwidth of Ca122 is $\sim$ 200 meV by the INS measurements and deduced the exchange parameters $SJ_{1a}$ = 49.9 meV, $SJ_{1b}$ = -5.7 meV, $SJ_{2}$ = 18.9 meV.~\cite{Zhao2009NP} These values are qualitatively in agreement with estimation by Han {\it et al.}~\cite{Han2009PRL} Also, the one-magnon energy can be approximately estimated by energy difference between the ordered state and the excited state with the one spin rotated toward the opposite direction. In the single-stripe AFM structure, the one-magnon energy can be estimated as $2S(J_{1a}-J_{1b}+2J_2)$ and this corresponds to 186.8 meV using the exchange parameters deduced by the INS measurements and roughly in agreement with the maximum one-magnon energy of the INS measurements. On the other hand, the two-magnon energy can be estimated by energy difference between the ordered state and the excited state with the two neighbor opposite spins rotated toward the each opposite directions, because the total spin quantum number should be conserved by the Raman scattering. Although there are two kinds of the two neighbor spins with the opposite directions in the single-stripe AFM structure, if we assume that the two-magnon excitation energy can be estimated by the minimum value of the energy difference between the two states, it can be estimated as $4S(J_{1a}-J_{1b}+2J_2)-J_{1a}$ and this corresponds to 323.7 meV by supposing $S = 1$. 

\begin{figure} [t]
\begin{center}
\includegraphics[width=8.5cm]{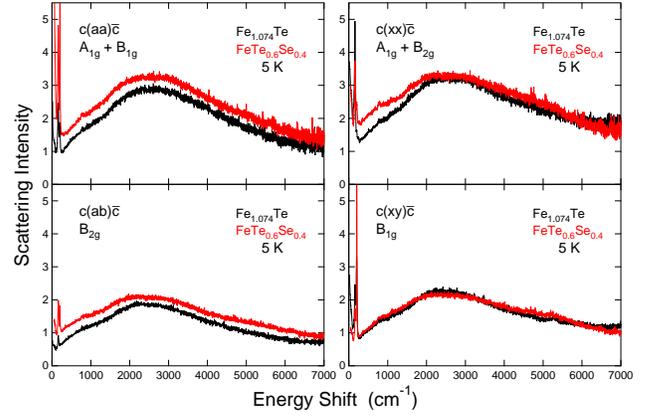} 
\caption{(Color online) Comparison of two-magnon spectra between the parent compound Fe$_{1.074}$Te and the superconducting FeTe$_{0.6}$Se$_{0.4}$.}
\label{Fig6} 
\end{center}
\end{figure}

For the case of 11 system, the exchange interactions have been estimated as $SJ_{1a}$ = -7.6 meV, $SJ_{1b}$ = -26.5 meV, $SJ_{2a}$ = 46.5 meV, and $SJ_{2b}$ = -34.9 meV for the double-stripe Fe$_{1.068}$Te by Han {\it et al.}~\cite{Han2009PRLa} based on the first-principle calculations. As the above approximate estimation, the one-magnon and two-magnon energy in the double-stripe AFM structure can be estimated as $2S(J_{1a}-J_{1b}+J_{2a}-J_{2b})$ and $4S(J_{1a}-J_{1b}+J_{2a}-J_{2b})-J_{2a}$, respectively, and these values correspond to 200.6 meV and 354.7 meV, respectively. The observed two-magnon peak is somewhat smaller than this estimated value. However, the peak energy of the two-magnon scattering can be smaller than the estimated value from the one-magnon energy because it depends on the details of magnon dispersion relations. Hence, we can say that the peak energy of the two-magnon scattering is reasonable and it is also reasonable that two-magnon energies are similar between Fe$_{1.074}$Te and Ba122. From the INS measurements of superconducting and non-superconducting FeTe$_{1-x}$Se$_x$, it is insisted that the spin excitation spectrum extends above 250 meV by Lumsden {\it et al.}.~\cite{Lumsden2010NP}  

Figure~\ref{Fig6} shows a comparison of the two-magnon spectra between the parent compound Fe$_{1.074}$Te and the superconducting FeTe$_{0.6}$Se$_{0.4}$. One can see that the two-magnon spectra are almost the same between these two compounds. This is rather surprising because FeTe$_{0.6}$Se$_{0.4}$ no longer shows long-range magnetic ordering. For the case of the hole-doped high-$T_c$ cuprates, while the two-magnon peak of the AF parent compounds is relatively sharp in the $B_{1g}$ spectra, it becomes broader and the peak energy becomes smaller with hole doping.~\cite{Sugai2003PRB} However, because Se substitution is isovalent doping and the parent compound Fe$_{1.074}$Te is an itinerant antiferromagnet, while the parent compound of the high-$T_c$ cuprates is an antiferromagnetic insulator, it may be reasonable that doping dependence is so different from the cuprates. Anyway from these results, we can say that almost the same magnetic excitations and/or magnetic fluctuations exist even in the superconducting FeTe$_{0.6}$Se$_{0.4}$ with the antiferromagnetic Fe$_{1.074}$Te. 

As for the selection rule of the two-magnon scattering, if the ordered magnetic moments are originated from the localized spins and Hamiltonian of the system can be described by the Heisenberg model including only the nearest neighbor exchange interactions, the two-magnon scattering Hamiltonian can be given by~\cite{Parkinson1969JPC}
\[
H^{2-mag}=A\sum_{<ij>} (\boldsymbol{E}_{inc} \cdot \boldsymbol{\sigma}_{ij})(\boldsymbol{E}_{sc} \cdot \boldsymbol{\sigma}_{ij})(\boldsymbol{S}_i \cdot \boldsymbol{S}_j),
\]
where $\boldsymbol{E}_{inc}$, $\boldsymbol{E}_{sc}$ are polarization vectors of the incident and scattered light, respectively, $\boldsymbol{\sigma}_{ij}$ is a unit vector connecting the nearest neighbor sites $i$ and $j$, and $\boldsymbol{S}_i$, $\boldsymbol{S}_j$ are spin operators at the site $i$ and $j$, respectively. Even using this Hamiltonian, the selection rule can be varied when the crystal and magnetic structures are different. In addition, when not only the nearest neighbor interactions but also further interactions exist, the selection rule can be complex. At least, when the second nearest neighbor interactions are finite, the two-magnon scattering should be active in $B_{1g}$ and $B_{2g}$ mode. In this case, two-magnon scattering can be observed in all our polarization configurations. However, for the case of the iron-based superconductors, since the electric resistivity of parent compounds are metallic, it should not have been settled whether the localized spins exist or not~\cite{Zhao2009NP}. Hence, the above selection rule might not be applied to the present system. Anyway, we consider that it would be intrinsic that the two-magnon peak has been observed for all our polarization configurations.

\begin{figure} [t]
\begin{center}
\includegraphics[width=8.5cm]{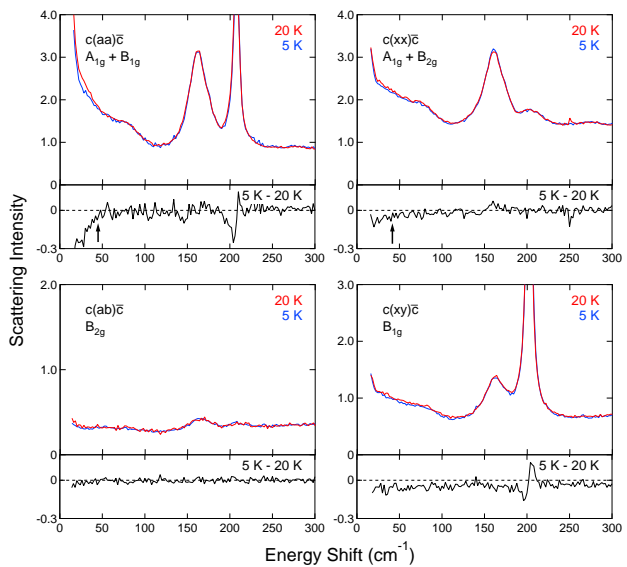}
\caption{(Color online) Electronic Raman spectra of FeTe$_{0.6}$Se$_{0.4}$ above and below $T_c$ and difference spectra. Dip structures are denoted by the arrows in the (aa) and (xx) configuration spectra.}
\label{Fig7} 
\end{center}
\end{figure}

\begin{figure*} [t]
\begin{center}
\includegraphics[width=18cm]{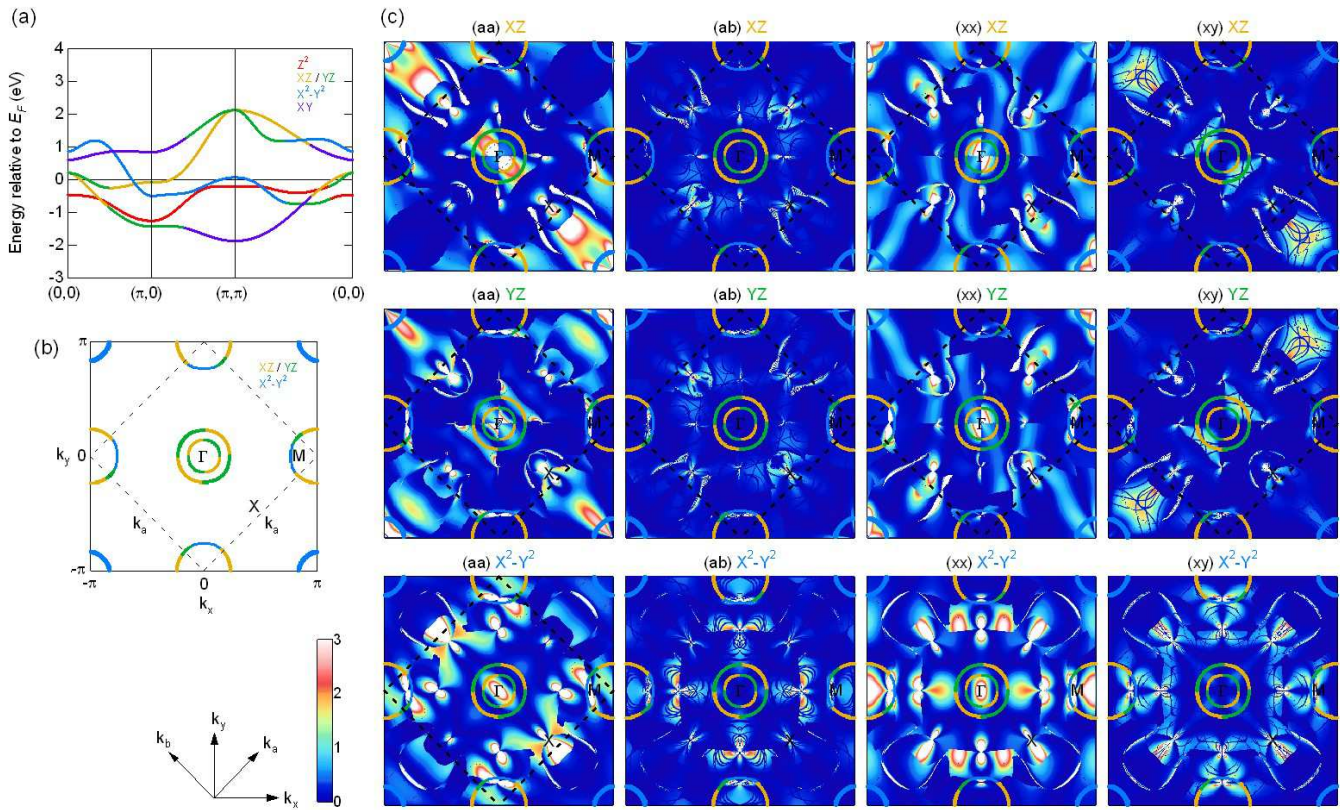} 
\caption{(Color online) The band structure of the unfolded 5-band tight-binding model reproduced from Ref.~\onlinecite{Kuroki2008PRL}. Colors are distinguished by the contributions of each orbital. Fermi level has shifted to reproduce three hole FS observed by ARPES.~\cite{Chen2010PRB} (b) Fermi surfaces deduced from the 5-band tight-binding model. (c) The top, middle, and bottom panels show the Raman form factors $| \frac{m}{\hbar^2} \partial ^2 \epsilon /\partial k_\alpha \partial k_\beta | ^2$ of all the measured polarization configurations for the XZ, YZ and X$^2$-Y$^2$ bands, respectively. The contributions to the electronic Raman scattering is significant at the white or red regions overlapped with each FS sheet represented by the yellow (XZ band), green (YZ band), and blue (X$^2$-Y$^2$ band) lines, respectively.}
\label{Fig8} 
\end{center}
\end{figure*}

\subsection{Electronic Raman Scattering}

Finally, we would like to show the spectra of the electronic Raman scattering. Figure~\ref{Fig7} shows the Raman spectra in the low-energy region measured below $T_c$ (5 K) and above $T_c$ (20 K), and the difference spectra between these two spectra. Scattering intensity below 100 cm$^{-1}$ should be contributed from the electronic scattering. Because the intensity of the electronic Raman scattering is proportional to the square of the inverse effective-mass tensor $| \frac{m}{\hbar^2} \partial ^2 \epsilon /\partial k_\alpha \partial k_\beta | ^2$,~\cite{Devereaux2007RMP} the momentum dependence of Raman scattering intensity can be complex for the case of the multiband system like iron-based superconductors. We have calculated the Raman form factor for each polarization configuration based on the unfolded 5-band tight-binding (TB) model. Figure~\ref{Fig7} (a) shows the band dispersions of the unfolded 5-band TB model reproduced from Ref.~\onlinecite{Kuroki2008PRL}. Fermi level ($E_F$) has been shifted to reproduce three hole FS observed by the ARPES measurements of Fe$_{1.04}$Te$_{0.66}$Se$_{0.34}$~\cite{Chen2010PRB} as shown in Fig.~\ref{Fig8} (b). Although the parameters of this TB model is evaluated from the first-principle band-structure calculation of LaFeAsO$_{1-x}$F$_x$, the curvatures of each band around $E_F$ are approximately the same as those of the first-principle calculations for FeTe.~\cite{Subedi2008PRB} This has been also confirmed by comparison with our own calculation results with the Quantum-Espresso code. The calculated momentum dependence of the electronic-Raman-scattering intensity is shown in Fig.~\ref{Fig8} (c). Because only the XZ, YZ and X$^2$-Y$^2$ bands cross the $E_F$, we have shown only these three bands. One can recognize that the electronic Raman scattering is intense for the (aa) and (xx) polarization, and that relatively large contributions are from the inner XZ/YZ hole FS around (0,0) point and the electronic FS around ($\pi$,0) point for both configurations. However, we must note that $A_{1g}$ spectra can be partially screened by the long-range Coulomb interactions.~\cite{Devereaux2007RMP} In the difference spectra shown in Fig~\ref{Fig7}, dip structures around 40-45 cm$^{-1}$ (= 5.0-5.6 meV) can be recognized for (aa) and (xx) configurations as denoted by the arrows. These dip structures can be attributed to decrease of density of states at $E_F$ below $T_c$. Although the pair-breaking peak has not been observed, which is probably because $T_c$ is too low to be observed by our measurements, if we assume that the dip positions corresponds to 2$\Delta$, this gives a reduced gap value 2$\Delta/k_BT_c$ = 4.0-4.5. These values are slightly larger than the reported value from the STS measurements,~\cite{Kato2009PRB} but our estimation of 2$\Delta$ should be only able to give the upper limit of this value. It is consistent with the $s_{\pm}$ symmetry of the superconducting gap that the dip structure can be observed in the (aa) and (xx) polarizations, because in these polarizations, the excitations of $A_{1g}$ symmetry can be observed. Hence, we have concluded that these temperature and polarization dependence of the electronic Raman spectra is consistent with the STM/STS results.~\cite{Kato2009PRB,Hanaguri2010Science}   

Recently, electronic Raman scattering of Ba(Fe$_{1-x}$Co$_x$)$_2$As$_2$ has been reported~\cite{Muschler2009PRB,Sugai2010PRB} and a clear pair breaking peak has been observed for $B_{2g}$ spectrum. Muschler {\it et al.} have insisted that the $B_{2g}$ spectrum mainly probes the electronic FS around M point, while the $A_{1g}$ spectrum has the largest contribution from the interband scattering involving the hole bands. On the other hand, Sugai {\it et al.} have proposed that the pairing symmetry of orbital combination is $B_{2g}$ and that of momentum space is $A_{1g}$. Mazin {\it et al.} have calculated the symmetry dependent electronic Raman scattering based on the first-principles calculations~\cite{Mazin2010PRB}. This difference between 122 system and 11 system would be originated from the difference of band structures. Mazin {\it et al.} have pointed out that the band structure of 122 system is more three-dimensional (3D) than other iron-based superconductors. They also have pointed out that because the crystallographic symmetry of 122 system is body-centered tetragonal (bct) and different from other iron-based superconductors, which have primitive tetragonal structures, band folding from the unfolded Brillouin zone is rather complex. Furthermore, because Ba(Fe$_{1-x}$Co$_x$)$_2$As$_2$ is an electron-doped system while FeTe$_{1-x}$Se$_{x}$ is an isovalent-doping system, the size of FSs also should be rather different. From these reasons, the superconducting gap may be observed in the different symmetries between Ba(Fe$_{1-x}$Co$_x$)$_2$As$_2$ and FeTe$_{1-x}$Se$_{x}$.   

\section{Conclusion}

In summary, we have observed the phonon, two-magnon, and electronic Raman scattering of the iron-based superconductor FeTe$_{1-x}$ Se$_{x}$. The $A_{1g}$ and $B_{1g}$ phonon modes at 5 K have been observed at 158 cm$^{-1}$ and 202 cm$^{-1}$, respectively, for the parent compound Fe$_{1.074}$Te and at 161 cm$^{-1}$ and 202 cm$^{-1}$, respectively, for the superconducting FeTe$_{0.6}$Se$_{0.4}$. These modes have been assigned to $c$-axis anti-phase vibration modes of the chalcogens and irons, respectively. The frequencies of the observed phonons are reasonably in accord with the calculated results. Two-magnon scattering has been observed as broad peak structures around 2300 cm$^{-1}$ ($\sim$ 285 meV) and the broad weak edge structure around 300-800 cm$^{-1}$ ($\sim$ 93 meV) has been observed below the magnetic and structural transition temperature $T_s$ (= 57 K) in the spectra of the parent compound. The broad weak edge structure observed below $T_s$ is considered as associated to the magnetic transition. We have estimated the two-magnon energy as $4S(J_{1a}-J_{1b}+J_{2a}-J_{2b})-J_{2a}$, which corresponds to 354.7 meV using the theoretically estimated values of exchange interactions. the observed two-magnon peak is somewhat smaller than the estimated value. However, because the peak energy of the two-magnon scattering depends on the details of magnon dispersion relations, we have concluded that the observed two-magnon excitation energy is reasonable. In the spectra of the superconducting compound FeTe$_{0.6}$Se$_{0.4}$, also, almost the same two-magnon spectra have been observed. This is rather surprising because FeTe$_{0.6}$Se$_{0.4}$ no longer has a long-range magnetic ordering. This indicates that almost the same magnetic excitations and/or magnetic fluctuations exist even in the superconducting FeTe$_{0.6}$Se$_{0.4}$ with the antiferromagnetic Fe$_{1.074}$Te. The difference of the electronic Raman spectra between below $T_c$ (5 K) and above $T_c$ (20 K) has been deduced and dip structures around 40-45 cm$^{-1}$ (= 5.0-5.6 meV) for (aa) and (xx) configurations. This means that the superconducting gap has a $A_{1g}$ symmetry and consistent with the STM/STS results.


\begin{thebibliography}{50}
\expandafter\ifx\csname natexlab\endcsname\relax\def\natexlab#1{#1}\fi
\expandafter\ifx\csname bibnamefont\endcsname\relax
  \def\bibnamefont#1{#1}\fi
\expandafter\ifx\csname bibfnamefont\endcsname\relax
  \def\bibfnamefont#1{#1}\fi
\expandafter\ifx\csname citenamefont\endcsname\relax
  \def\citenamefont#1{#1}\fi
\expandafter\ifx\csname url\endcsname\relax
  \def\url#1{\texttt{#1}}\fi
\expandafter\ifx\csname urlprefix\endcsname\relax\def\urlprefix{URL }\fi
\providecommand{\bibinfo}[2]{#2}
\providecommand{\eprint}[2][]{\url{#2}}

\bibitem[{\citenamefont{Kamihara et~al.}(2008)\citenamefont{Kamihara, Watanabe,
  Hirano, and Hosono}}]{Kamihara2008JACS}
\bibinfo{author}{\bibfnamefont{Y.}~\bibnamefont{Kamihara}},
  \bibinfo{author}{\bibfnamefont{T.}~\bibnamefont{Watanabe}},
  \bibinfo{author}{\bibfnamefont{M.}~\bibnamefont{Hirano}}, \bibnamefont{and}
  \bibinfo{author}{\bibfnamefont{H.}~\bibnamefont{Hosono}},
  \bibinfo{journal}{J. Am. Chem. Soc} \textbf{\bibinfo{volume}{130}},
  \bibinfo{pages}{3296} (\bibinfo{year}{2008}).

\bibitem[{\citenamefont{Takahashi et~al.}(2008)\citenamefont{Takahashi, Igawa,
  Arii, Kamihara, Hirano, and Hosono}}]{Takahashi2008Nature}
\bibinfo{author}{\bibfnamefont{H.}~\bibnamefont{Takahashi}},
  \bibinfo{author}{\bibfnamefont{K.}~\bibnamefont{Igawa}},
  \bibinfo{author}{\bibfnamefont{K.}~\bibnamefont{Arii}},
  \bibinfo{author}{\bibfnamefont{Y.}~\bibnamefont{Kamihara}},
  \bibinfo{author}{\bibfnamefont{M.}~\bibnamefont{Hirano}}, \bibnamefont{and}
  \bibinfo{author}{\bibfnamefont{H.}~\bibnamefont{Hosono}},
  \bibinfo{journal}{Nature} \textbf{\bibinfo{volume}{453}},
  \bibinfo{pages}{376} (\bibinfo{year}{2008}), ISSN \bibinfo{issn}{0028-0836}.

\bibitem[{\citenamefont{Ren et~al.}(2008)\citenamefont{Ren, Yang, Lu, Yi, Shen,
  Li, Che, Dong, Sun, Zhou et~al.}}]{Ren2008CPL}
\bibinfo{author}{\bibfnamefont{Z.-A.} \bibnamefont{Ren}},
  \bibinfo{author}{\bibfnamefont{J.}~\bibnamefont{Yang}},
  \bibinfo{author}{\bibfnamefont{W.}~\bibnamefont{Lu}},
  \bibinfo{author}{\bibfnamefont{W.}~\bibnamefont{Yi}},
  \bibinfo{author}{\bibfnamefont{X.-L.} \bibnamefont{Shen}},
  \bibinfo{author}{\bibfnamefont{Z.-C.} \bibnamefont{Li}},
  \bibinfo{author}{\bibfnamefont{G.-C.} \bibnamefont{Che}},
  \bibinfo{author}{\bibfnamefont{X.-L.} \bibnamefont{Dong}},
  \bibinfo{author}{\bibfnamefont{L.-L.} \bibnamefont{Sun}},
  \bibinfo{author}{\bibfnamefont{F.}~\bibnamefont{Zhou}}, \bibnamefont{et~al.},
  \bibinfo{journal}{Chinese Physics Letters} \textbf{\bibinfo{volume}{25}},
  \bibinfo{pages}{2215} (\bibinfo{year}{2008}).

\bibitem[{\citenamefont{Rotter et~al.}(2008{\natexlab{a}})\citenamefont{Rotter,
  Tegel, Johrendt, Schellenberg, Hermes, and P\"ottgen}}]{Rotter2008PRB}
\bibinfo{author}{\bibfnamefont{M.}~\bibnamefont{Rotter}},
  \bibinfo{author}{\bibfnamefont{M.}~\bibnamefont{Tegel}},
  \bibinfo{author}{\bibfnamefont{D.}~\bibnamefont{Johrendt}},
  \bibinfo{author}{\bibfnamefont{I.}~\bibnamefont{Schellenberg}},
  \bibinfo{author}{\bibfnamefont{W.}~\bibnamefont{Hermes}}, \bibnamefont{and}
  \bibinfo{author}{\bibfnamefont{R.}~\bibnamefont{P\"ottgen}},
  \bibinfo{journal}{Phys. Rev. B} \textbf{\bibinfo{volume}{78}},
  \bibinfo{pages}{020503} (\bibinfo{year}{2008}{\natexlab{a}}).

\bibitem[{\citenamefont{Rotter et~al.}(2008{\natexlab{b}})\citenamefont{Rotter,
  Tegel, and Johrendt}}]{Rotter2008PRL}
\bibinfo{author}{\bibfnamefont{M.}~\bibnamefont{Rotter}},
  \bibinfo{author}{\bibfnamefont{M.}~\bibnamefont{Tegel}}, \bibnamefont{and}
  \bibinfo{author}{\bibfnamefont{D.}~\bibnamefont{Johrendt}},
  \bibinfo{journal}{Phys. Rev. Lett.} \textbf{\bibinfo{volume}{101}},
  \bibinfo{pages}{107006} (\bibinfo{year}{2008}{\natexlab{b}}).

\bibitem[{\citenamefont{Tapp et~al.}(2008)\citenamefont{Tapp, Tang, Lv, Sasmal,
  Lorenz, Chu, and Guloy}}]{Tapp2008PRB}
\bibinfo{author}{\bibfnamefont{J.~H.} \bibnamefont{Tapp}},
  \bibinfo{author}{\bibfnamefont{Z.}~\bibnamefont{Tang}},
  \bibinfo{author}{\bibfnamefont{B.}~\bibnamefont{Lv}},
  \bibinfo{author}{\bibfnamefont{K.}~\bibnamefont{Sasmal}},
  \bibinfo{author}{\bibfnamefont{B.}~\bibnamefont{Lorenz}},
  \bibinfo{author}{\bibfnamefont{P.~C.~W.} \bibnamefont{Chu}},
  \bibnamefont{and} \bibinfo{author}{\bibfnamefont{A.~M.} \bibnamefont{Guloy}},
  \bibinfo{journal}{Phys. Rev. B} \textbf{\bibinfo{volume}{78}},
  \bibinfo{pages}{060505} (\bibinfo{year}{2008}).

\bibitem[{\citenamefont{Wang et~al.}(2008)\citenamefont{Wang, Liu, Lv, Gao,
  Yang, Yu, Li, and Jin}}]{Wang2008SSC}
\bibinfo{author}{\bibfnamefont{X.}~\bibnamefont{Wang}},
  \bibinfo{author}{\bibfnamefont{Q.}~\bibnamefont{Liu}},
  \bibinfo{author}{\bibfnamefont{Y.}~\bibnamefont{Lv}},
  \bibinfo{author}{\bibfnamefont{W.}~\bibnamefont{Gao}},
  \bibinfo{author}{\bibfnamefont{L.}~\bibnamefont{Yang}},
  \bibinfo{author}{\bibfnamefont{R.}~\bibnamefont{Yu}},
  \bibinfo{author}{\bibfnamefont{F.}~\bibnamefont{Li}}, \bibnamefont{and}
  \bibinfo{author}{\bibfnamefont{C.}~\bibnamefont{Jin}},
  \bibinfo{journal}{Solid State Communications} \textbf{\bibinfo{volume}{148}},
  \bibinfo{pages}{538 } (\bibinfo{year}{2008}).

\bibitem[{\citenamefont{Hsu et~al.}(2008)\citenamefont{Hsu, Luo, Yeh, Chen,
  Huang, Wu, Lee, Huang, Chu, Yan et~al.}}]{Hsu2008PNAS}
\bibinfo{author}{\bibfnamefont{F.-C.} \bibnamefont{Hsu}},
  \bibinfo{author}{\bibfnamefont{J.-Y.} \bibnamefont{Luo}},
  \bibinfo{author}{\bibfnamefont{K.-W.} \bibnamefont{Yeh}},
  \bibinfo{author}{\bibfnamefont{T.-K.} \bibnamefont{Chen}},
  \bibinfo{author}{\bibfnamefont{T.-W.} \bibnamefont{Huang}},
  \bibinfo{author}{\bibfnamefont{P.~M.} \bibnamefont{Wu}},
  \bibinfo{author}{\bibfnamefont{Y.-C.} \bibnamefont{Lee}},
  \bibinfo{author}{\bibfnamefont{Y.-L.} \bibnamefont{Huang}},
  \bibinfo{author}{\bibfnamefont{Y.-Y.} \bibnamefont{Chu}},
  \bibinfo{author}{\bibfnamefont{D.-C.} \bibnamefont{Yan}},
  \bibnamefont{et~al.}, \bibinfo{journal}{Proc. Natl. Acad. Sci. U.S.A.}
  \textbf{\bibinfo{volume}{105}}, \bibinfo{pages}{14262}
  (\bibinfo{year}{2008}).

\bibitem[{\citenamefont{Yeh et~al.}(2008)\citenamefont{Yeh, Huang, lin Huang,
  Chen, Hsu, Wu, Lee, Chu, Chen, Luo et~al.}}]{Yeh2008EPL}
\bibinfo{author}{\bibfnamefont{K.-W.} \bibnamefont{Yeh}},
  \bibinfo{author}{\bibfnamefont{T.-W.} \bibnamefont{Huang}},
  \bibinfo{author}{\bibfnamefont{Y.}~\bibnamefont{lin Huang}},
  \bibinfo{author}{\bibfnamefont{T.-K.} \bibnamefont{Chen}},
  \bibinfo{author}{\bibfnamefont{F.-C.} \bibnamefont{Hsu}},
  \bibinfo{author}{\bibfnamefont{P.~M.} \bibnamefont{Wu}},
  \bibinfo{author}{\bibfnamefont{Y.-C.} \bibnamefont{Lee}},
  \bibinfo{author}{\bibfnamefont{Y.-Y.} \bibnamefont{Chu}},
  \bibinfo{author}{\bibfnamefont{C.-L.} \bibnamefont{Chen}},
  \bibinfo{author}{\bibfnamefont{J.-Y.} \bibnamefont{Luo}},
  \bibnamefont{et~al.}, \bibinfo{journal}{Europhys. Lett.}
  \textbf{\bibinfo{volume}{84}}, \bibinfo{pages}{37002} (\bibinfo{year}{2008}).

\bibitem[{\citenamefont{Fang et~al.}(2008)\citenamefont{Fang, Pham, Qian, Liu,
  Vehstedt, Liu, Spinu, and Mao}}]{Fang2008PRB}
\bibinfo{author}{\bibfnamefont{M.~H.} \bibnamefont{Fang}},
  \bibinfo{author}{\bibfnamefont{H.~M.} \bibnamefont{Pham}},
  \bibinfo{author}{\bibfnamefont{B.}~\bibnamefont{Qian}},
  \bibinfo{author}{\bibfnamefont{T.~J.} \bibnamefont{Liu}},
  \bibinfo{author}{\bibfnamefont{E.~K.} \bibnamefont{Vehstedt}},
  \bibinfo{author}{\bibfnamefont{Y.}~\bibnamefont{Liu}},
  \bibinfo{author}{\bibfnamefont{L.}~\bibnamefont{Spinu}}, \bibnamefont{and}
  \bibinfo{author}{\bibfnamefont{Z.~Q.} \bibnamefont{Mao}},
  \bibinfo{journal}{Phys. Rev. B} \textbf{\bibinfo{volume}{78}},
  \bibinfo{pages}{224503} (\bibinfo{year}{2008}).

\bibitem[{\citenamefont{Zhu et~al.}(2009)\citenamefont{Zhu, Han, Mu, Cheng,
  Shen, Zeng, and Wen}}]{Zhu2009PRB}
\bibinfo{author}{\bibfnamefont{X.}~\bibnamefont{Zhu}},
  \bibinfo{author}{\bibfnamefont{F.}~\bibnamefont{Han}},
  \bibinfo{author}{\bibfnamefont{G.}~\bibnamefont{Mu}},
  \bibinfo{author}{\bibfnamefont{P.}~\bibnamefont{Cheng}},
  \bibinfo{author}{\bibfnamefont{B.}~\bibnamefont{Shen}},
  \bibinfo{author}{\bibfnamefont{B.}~\bibnamefont{Zeng}}, \bibnamefont{and}
  \bibinfo{author}{\bibfnamefont{H.-H.} \bibnamefont{Wen}},
  \bibinfo{journal}{Phys. Rev. B} \textbf{\bibinfo{volume}{79}},
  \bibinfo{pages}{220512} (\bibinfo{year}{2009}).

\bibitem[{\citenamefont{Singh and Du}(2008)}]{Singh2008PRL}
\bibinfo{author}{\bibfnamefont{D.~J.} \bibnamefont{Singh}} \bibnamefont{and}
  \bibinfo{author}{\bibfnamefont{M.-H.} \bibnamefont{Du}},
  \bibinfo{journal}{Phys. Rev. Lett.} \textbf{\bibinfo{volume}{100}},
  \bibinfo{pages}{237003} (\bibinfo{year}{2008}).

\bibitem[{\citenamefont{Singh}(2008)}]{Singh2008PRB}
\bibinfo{author}{\bibfnamefont{D.~J.} \bibnamefont{Singh}},
  \bibinfo{journal}{Phys. Rev. B} \textbf{\bibinfo{volume}{78}},
  \bibinfo{pages}{094511} (\bibinfo{year}{2008}).

\bibitem[{\citenamefont{Subedi et~al.}(2008)\citenamefont{Subedi, Zhang, Singh,
  and Du}}]{Subedi2008PRB}
\bibinfo{author}{\bibfnamefont{A.}~\bibnamefont{Subedi}},
  \bibinfo{author}{\bibfnamefont{L.}~\bibnamefont{Zhang}},
  \bibinfo{author}{\bibfnamefont{D.~J.} \bibnamefont{Singh}}, \bibnamefont{and}
  \bibinfo{author}{\bibfnamefont{M.~H.} \bibnamefont{Du}},
  \bibinfo{journal}{Phys. Rev. B} \textbf{\bibinfo{volume}{78}},
  \bibinfo{pages}{134514} (\bibinfo{year}{2008}).

\bibitem[{\citenamefont{Wang et~al.}(2009)\citenamefont{Wang, Zhang, Zheng, and
  Yang}}]{Wang2009PRB}
\bibinfo{author}{\bibfnamefont{G.}~\bibnamefont{Wang}},
  \bibinfo{author}{\bibfnamefont{M.}~\bibnamefont{Zhang}},
  \bibinfo{author}{\bibfnamefont{L.}~\bibnamefont{Zheng}}, \bibnamefont{and}
  \bibinfo{author}{\bibfnamefont{Z.}~\bibnamefont{Yang}},
  \bibinfo{journal}{Phys. Rev. B} \textbf{\bibinfo{volume}{80}},
  \bibinfo{pages}{184501} (\bibinfo{year}{2009}).

\bibitem[{\citenamefont{Dong et~al.}(2008)\citenamefont{Dong, Zhang, Xu, Li,
  Li, Hu, Wu, Chen, Dai, Luo et~al.}}]{Dong2008EPL}
\bibinfo{author}{\bibfnamefont{J.}~\bibnamefont{Dong}},
  \bibinfo{author}{\bibfnamefont{H.~J.} \bibnamefont{Zhang}},
  \bibinfo{author}{\bibfnamefont{G.}~\bibnamefont{Xu}},
  \bibinfo{author}{\bibfnamefont{Z.}~\bibnamefont{Li}},
  \bibinfo{author}{\bibfnamefont{G.}~\bibnamefont{Li}},
  \bibinfo{author}{\bibfnamefont{W.~Z.} \bibnamefont{Hu}},
  \bibinfo{author}{\bibfnamefont{D.}~\bibnamefont{Wu}},
  \bibinfo{author}{\bibfnamefont{G.~F.} \bibnamefont{Chen}},
  \bibinfo{author}{\bibfnamefont{X.}~\bibnamefont{Dai}},
  \bibinfo{author}{\bibfnamefont{J.~L.} \bibnamefont{Luo}},
  \bibnamefont{et~al.}, \bibinfo{journal}{Europhysics Letters}
  \textbf{\bibinfo{volume}{83}}, \bibinfo{pages}{27006} (\bibinfo{year}{2008}).

\bibitem[{\citenamefont{Xu et~al.}(2008)\citenamefont{Xu, Zhang, Dai, and
  Fang}}]{Xu2008EPL}
\bibinfo{author}{\bibfnamefont{G.}~\bibnamefont{Xu}},
  \bibinfo{author}{\bibfnamefont{H.}~\bibnamefont{Zhang}},
  \bibinfo{author}{\bibfnamefont{X.}~\bibnamefont{Dai}}, \bibnamefont{and}
  \bibinfo{author}{\bibfnamefont{Z.}~\bibnamefont{Fang}},
  \bibinfo{journal}{Europhysics Letters} \textbf{\bibinfo{volume}{84}},
  \bibinfo{pages}{67015} (\bibinfo{year}{2008}).

\bibitem[{\citenamefont{Deng et~al.}(2009)\citenamefont{Deng, K\"ohler, and
  Simon}}]{Deng2009PRB}
\bibinfo{author}{\bibfnamefont{S.}~\bibnamefont{Deng}},
  \bibinfo{author}{\bibfnamefont{J.}~\bibnamefont{K\"ohler}}, \bibnamefont{and}
  \bibinfo{author}{\bibfnamefont{A.}~\bibnamefont{Simon}},
  \bibinfo{journal}{Phys. Rev. B} \textbf{\bibinfo{volume}{80}},
  \bibinfo{pages}{214508} (\bibinfo{year}{2009}).

\bibitem[{\citenamefont{Mazin et~al.}(2008)\citenamefont{Mazin, Singh,
  Johannes, and Du}}]{Mazin2008PRL}
\bibinfo{author}{\bibfnamefont{I.~I.} \bibnamefont{Mazin}},
  \bibinfo{author}{\bibfnamefont{D.~J.} \bibnamefont{Singh}},
  \bibinfo{author}{\bibfnamefont{M.~D.} \bibnamefont{Johannes}},
  \bibnamefont{and} \bibinfo{author}{\bibfnamefont{M.~H.} \bibnamefont{Du}},
  \bibinfo{journal}{Phys. Rev. Lett.} \textbf{\bibinfo{volume}{101}},
  \bibinfo{pages}{057003} (\bibinfo{year}{2008}).

\bibitem[{\citenamefont{Kuroki et~al.}(2008)\citenamefont{Kuroki, Onari, Arita,
  Usui, Tanaka, Kontani, and Aoki}}]{Kuroki2008PRL}
\bibinfo{author}{\bibfnamefont{K.}~\bibnamefont{Kuroki}},
  \bibinfo{author}{\bibfnamefont{S.}~\bibnamefont{Onari}},
  \bibinfo{author}{\bibfnamefont{R.}~\bibnamefont{Arita}},
  \bibinfo{author}{\bibfnamefont{H.}~\bibnamefont{Usui}},
  \bibinfo{author}{\bibfnamefont{Y.}~\bibnamefont{Tanaka}},
  \bibinfo{author}{\bibfnamefont{H.}~\bibnamefont{Kontani}}, \bibnamefont{and}
  \bibinfo{author}{\bibfnamefont{H.}~\bibnamefont{Aoki}},
  \bibinfo{journal}{Phys. Rev. Lett.} \textbf{\bibinfo{volume}{101}},
  \bibinfo{pages}{087004} (\bibinfo{year}{2008}).

\bibitem[{\citenamefont{Ding et~al.}(2008)\citenamefont{Ding, Richard,
  Nakayama, Sugawara, Arakane, Sekiba, Takayama, Souma, Sato, Takahashi
  et~al.}}]{Ding2008EPL}
\bibinfo{author}{\bibfnamefont{H.}~\bibnamefont{Ding}},
  \bibinfo{author}{\bibfnamefont{P.}~\bibnamefont{Richard}},
  \bibinfo{author}{\bibfnamefont{K.}~\bibnamefont{Nakayama}},
  \bibinfo{author}{\bibfnamefont{K.}~\bibnamefont{Sugawara}},
  \bibinfo{author}{\bibfnamefont{T.}~\bibnamefont{Arakane}},
  \bibinfo{author}{\bibfnamefont{Y.}~\bibnamefont{Sekiba}},
  \bibinfo{author}{\bibfnamefont{A.}~\bibnamefont{Takayama}},
  \bibinfo{author}{\bibfnamefont{S.}~\bibnamefont{Souma}},
  \bibinfo{author}{\bibfnamefont{T.}~\bibnamefont{Sato}},
  \bibinfo{author}{\bibfnamefont{T.}~\bibnamefont{Takahashi}},
  \bibnamefont{et~al.}, \bibinfo{journal}{Europhysics Letters}
  \textbf{\bibinfo{volume}{83}}, \bibinfo{pages}{47001} (\bibinfo{year}{2008}).

\bibitem[{\citenamefont{Xia et~al.}(2009{\natexlab{a}})\citenamefont{Xia, Qian,
  Wray, Hsieh, Chen, Luo, Wang, and Hasan}}]{Xia2009PRL}
\bibinfo{author}{\bibfnamefont{Y.}~\bibnamefont{Xia}},
  \bibinfo{author}{\bibfnamefont{D.}~\bibnamefont{Qian}},
  \bibinfo{author}{\bibfnamefont{L.}~\bibnamefont{Wray}},
  \bibinfo{author}{\bibfnamefont{D.}~\bibnamefont{Hsieh}},
  \bibinfo{author}{\bibfnamefont{G.~F.} \bibnamefont{Chen}},
  \bibinfo{author}{\bibfnamefont{J.~L.} \bibnamefont{Luo}},
  \bibinfo{author}{\bibfnamefont{N.~L.} \bibnamefont{Wang}}, \bibnamefont{and}
  \bibinfo{author}{\bibfnamefont{M.~Z.} \bibnamefont{Hasan}},
  \bibinfo{journal}{Phys. Rev. Lett.} \textbf{\bibinfo{volume}{103}},
  \bibinfo{pages}{037002} (\bibinfo{year}{2009}{\natexlab{a}}).

\bibitem[{\citenamefont{de~la Cruz et~al.}(2008)\citenamefont{de~la Cruz,
  Huang, Lynn, Li, II, Zarestky, Mook, Chen, Luo, Wang
  et~al.}}]{Cruz2008Nature}
\bibinfo{author}{\bibfnamefont{C.}~\bibnamefont{de~la Cruz}},
  \bibinfo{author}{\bibfnamefont{Q.}~\bibnamefont{Huang}},
  \bibinfo{author}{\bibfnamefont{J.~W.} \bibnamefont{Lynn}},
  \bibinfo{author}{\bibfnamefont{J.}~\bibnamefont{Li}},
  \bibinfo{author}{\bibfnamefont{W.~R.} \bibnamefont{II}},
  \bibinfo{author}{\bibfnamefont{J.~L.} \bibnamefont{Zarestky}},
  \bibinfo{author}{\bibfnamefont{H.~A.} \bibnamefont{Mook}},
  \bibinfo{author}{\bibfnamefont{G.~F.} \bibnamefont{Chen}},
  \bibinfo{author}{\bibfnamefont{J.~L.} \bibnamefont{Luo}},
  \bibinfo{author}{\bibfnamefont{N.~L.} \bibnamefont{Wang}},
  \bibnamefont{et~al.}, \bibinfo{journal}{Nature}
  \textbf{\bibinfo{volume}{453}}, \bibinfo{pages}{899} (\bibinfo{year}{2008}).

\bibitem[{\citenamefont{Huang et~al.}(2008)\citenamefont{Huang, Qiu, Bao,
  Green, Lynn, Gasparovic, Wu, Wu, and Chen}}]{Huang2008PRL}
\bibinfo{author}{\bibfnamefont{Q.}~\bibnamefont{Huang}},
  \bibinfo{author}{\bibfnamefont{Y.}~\bibnamefont{Qiu}},
  \bibinfo{author}{\bibfnamefont{W.}~\bibnamefont{Bao}},
  \bibinfo{author}{\bibfnamefont{M.~A.} \bibnamefont{Green}},
  \bibinfo{author}{\bibfnamefont{J.~W.} \bibnamefont{Lynn}},
  \bibinfo{author}{\bibfnamefont{Y.~C.} \bibnamefont{Gasparovic}},
  \bibinfo{author}{\bibfnamefont{T.}~\bibnamefont{Wu}},
  \bibinfo{author}{\bibfnamefont{G.}~\bibnamefont{Wu}}, \bibnamefont{and}
  \bibinfo{author}{\bibfnamefont{X.~H.} \bibnamefont{Chen}},
  \bibinfo{journal}{Phys. Rev. Lett.} \textbf{\bibinfo{volume}{101}},
  \bibinfo{pages}{257003} (\bibinfo{year}{2008}).

\bibitem[{\citenamefont{Li et~al.}(2009)\citenamefont{Li, de~la Cruz, Huang,
  Chen, Lynn, Hu, Huang, Hsu, Yeh, Wu et~al.}}]{Li2009PRB}
\bibinfo{author}{\bibfnamefont{S.}~\bibnamefont{Li}},
  \bibinfo{author}{\bibfnamefont{C.}~\bibnamefont{de~la Cruz}},
  \bibinfo{author}{\bibfnamefont{Q.}~\bibnamefont{Huang}},
  \bibinfo{author}{\bibfnamefont{Y.}~\bibnamefont{Chen}},
  \bibinfo{author}{\bibfnamefont{J.~W.} \bibnamefont{Lynn}},
  \bibinfo{author}{\bibfnamefont{J.}~\bibnamefont{Hu}},
  \bibinfo{author}{\bibfnamefont{Y.-L.} \bibnamefont{Huang}},
  \bibinfo{author}{\bibfnamefont{F.-C.} \bibnamefont{Hsu}},
  \bibinfo{author}{\bibfnamefont{K.-W.} \bibnamefont{Yeh}},
  \bibinfo{author}{\bibfnamefont{M.-K.} \bibnamefont{Wu}},
  \bibnamefont{et~al.}, \bibinfo{journal}{Phys. Rev. B}
  \textbf{\bibinfo{volume}{79}}, \bibinfo{pages}{054503}
  (\bibinfo{year}{2009}).

\bibitem[{\citenamefont{Medvedev et~al.}(2009)\citenamefont{Medvedev, McQueen,
  Troyan, Palasyuk, Eremets, Cava, Naghavi, Casper, Ksenofontov, Wortmann
  et~al.}}]{Medvedev2009NM}
\bibinfo{author}{\bibfnamefont{S.}~\bibnamefont{Medvedev}},
  \bibinfo{author}{\bibfnamefont{T.~M.} \bibnamefont{McQueen}},
  \bibinfo{author}{\bibfnamefont{I.~A.} \bibnamefont{Troyan}},
  \bibinfo{author}{\bibfnamefont{T.}~\bibnamefont{Palasyuk}},
  \bibinfo{author}{\bibfnamefont{M.~I.} \bibnamefont{Eremets}},
  \bibinfo{author}{\bibfnamefont{R.~J.} \bibnamefont{Cava}},
  \bibinfo{author}{\bibfnamefont{S.}~\bibnamefont{Naghavi}},
  \bibinfo{author}{\bibfnamefont{F.}~\bibnamefont{Casper}},
  \bibinfo{author}{\bibfnamefont{V.}~\bibnamefont{Ksenofontov}},
  \bibinfo{author}{\bibfnamefont{G.}~\bibnamefont{Wortmann}},
  \bibnamefont{et~al.}, \bibinfo{journal}{Nature Mater}
  \textbf{\bibinfo{volume}{8}}, \bibinfo{pages}{630} (\bibinfo{year}{2009}).

\bibitem[{\citenamefont{Margadonna et~al.}(2009)\citenamefont{Margadonna,
  Takabayashi, Ohishi, Mizuguchi, Takano, Kagayama, Nakagawa, Takata, and
  Prassides}}]{Margadonna2009PRB}
\bibinfo{author}{\bibfnamefont{S.}~\bibnamefont{Margadonna}},
  \bibinfo{author}{\bibfnamefont{Y.}~\bibnamefont{Takabayashi}},
  \bibinfo{author}{\bibfnamefont{Y.}~\bibnamefont{Ohishi}},
  \bibinfo{author}{\bibfnamefont{Y.}~\bibnamefont{Mizuguchi}},
  \bibinfo{author}{\bibfnamefont{Y.}~\bibnamefont{Takano}},
  \bibinfo{author}{\bibfnamefont{T.}~\bibnamefont{Kagayama}},
  \bibinfo{author}{\bibfnamefont{T.}~\bibnamefont{Nakagawa}},
  \bibinfo{author}{\bibfnamefont{M.}~\bibnamefont{Takata}}, \bibnamefont{and}
  \bibinfo{author}{\bibfnamefont{K.}~\bibnamefont{Prassides}},
  \bibinfo{journal}{Phys. Rev. B} \textbf{\bibinfo{volume}{80}},
  \bibinfo{pages}{064506} (\bibinfo{year}{2009}).

\bibitem[{\citenamefont{Margadonna et~al.}(2008)\citenamefont{Margadonna,
  Takabayashi, McDonald, Kasperkiewicz, Mizuguchi, Takano, Fitch, Suard, and
  Prassides}}]{Margadonna2008CC}
\bibinfo{author}{\bibfnamefont{S.}~\bibnamefont{Margadonna}},
  \bibinfo{author}{\bibfnamefont{Y.}~\bibnamefont{Takabayashi}},
  \bibinfo{author}{\bibfnamefont{M.~T.} \bibnamefont{McDonald}},
  \bibinfo{author}{\bibfnamefont{K.}~\bibnamefont{Kasperkiewicz}},
  \bibinfo{author}{\bibfnamefont{Y.}~\bibnamefont{Mizuguchi}},
  \bibinfo{author}{\bibfnamefont{Y.}~\bibnamefont{Takano}},
  \bibinfo{author}{\bibfnamefont{A.~N.} \bibnamefont{Fitch}},
  \bibinfo{author}{\bibfnamefont{E.}~\bibnamefont{Suard}}, \bibnamefont{and}
  \bibinfo{author}{\bibfnamefont{K.}~\bibnamefont{Prassides}},
  \bibinfo{journal}{Chem. Commun.} \textbf{\bibinfo{volume}{2008}},
  \bibinfo{pages}{5607} (\bibinfo{year}{2008}).

\bibitem[{\citenamefont{Bao et~al.}(2009)\citenamefont{Bao, Qiu, Huang, Green,
  Zajdel, Fitzsimmons, Zhernenkov, Chang, Fang, Qian et~al.}}]{Bao2009PRL}
\bibinfo{author}{\bibfnamefont{W.}~\bibnamefont{Bao}},
  \bibinfo{author}{\bibfnamefont{Y.}~\bibnamefont{Qiu}},
  \bibinfo{author}{\bibfnamefont{Q.}~\bibnamefont{Huang}},
  \bibinfo{author}{\bibfnamefont{M.~A.} \bibnamefont{Green}},
  \bibinfo{author}{\bibfnamefont{P.}~\bibnamefont{Zajdel}},
  \bibinfo{author}{\bibfnamefont{M.~R.} \bibnamefont{Fitzsimmons}},
  \bibinfo{author}{\bibfnamefont{M.}~\bibnamefont{Zhernenkov}},
  \bibinfo{author}{\bibfnamefont{S.}~\bibnamefont{Chang}},
  \bibinfo{author}{\bibfnamefont{M.}~\bibnamefont{Fang}},
  \bibinfo{author}{\bibfnamefont{B.}~\bibnamefont{Qian}}, \bibnamefont{et~al.},
  \bibinfo{journal}{Phys. Rev. Lett.} \textbf{\bibinfo{volume}{102}},
  \bibinfo{pages}{247001} (\bibinfo{year}{2009}).

\bibitem[{\citenamefont{Qiu et~al.}(2009)\citenamefont{Qiu, Bao, Zhao, Broholm,
  Stanev, Tesanovic, Gasparovic, Chang, Hu, Qian et~al.}}]{Qiu2009PRL}
\bibinfo{author}{\bibfnamefont{Y.}~\bibnamefont{Qiu}},
  \bibinfo{author}{\bibfnamefont{W.}~\bibnamefont{Bao}},
  \bibinfo{author}{\bibfnamefont{Y.}~\bibnamefont{Zhao}},
  \bibinfo{author}{\bibfnamefont{C.}~\bibnamefont{Broholm}},
  \bibinfo{author}{\bibfnamefont{V.}~\bibnamefont{Stanev}},
  \bibinfo{author}{\bibfnamefont{Z.}~\bibnamefont{Tesanovic}},
  \bibinfo{author}{\bibfnamefont{Y.~C.} \bibnamefont{Gasparovic}},
  \bibinfo{author}{\bibfnamefont{S.}~\bibnamefont{Chang}},
  \bibinfo{author}{\bibfnamefont{J.}~\bibnamefont{Hu}},
  \bibinfo{author}{\bibfnamefont{B.}~\bibnamefont{Qian}}, \bibnamefont{et~al.},
  \bibinfo{journal}{Phys. Rev. Lett.} \textbf{\bibinfo{volume}{103}},
  \bibinfo{pages}{067008} (\bibinfo{year}{2009}).

\bibitem[{\citenamefont{Hanaguri et~al.}(2010)\citenamefont{Hanaguri, Niitaka,
  Kuroki, and Takagi}}]{Hanaguri2010Science}
\bibinfo{author}{\bibfnamefont{T.}~\bibnamefont{Hanaguri}},
  \bibinfo{author}{\bibfnamefont{S.}~\bibnamefont{Niitaka}},
  \bibinfo{author}{\bibfnamefont{K.}~\bibnamefont{Kuroki}}, \bibnamefont{and}
  \bibinfo{author}{\bibfnamefont{H.}~\bibnamefont{Takagi}},
  \bibinfo{journal}{Science} \textbf{\bibinfo{volume}{328}},
  \bibinfo{pages}{474} (\bibinfo{year}{2010}).

\bibitem[{\citenamefont{Giannozzi et~al.}(2009)\citenamefont{Giannozzi, Baroni,
  Bonini, Calandra, Car, Cavazzoni, Ceresoli, Chiarotti, Cococcioni, Dabo
  et~al.}}]{QE}
\bibinfo{author}{\bibfnamefont{P.}~\bibnamefont{Giannozzi}},
  \bibinfo{author}{\bibfnamefont{S.}~\bibnamefont{Baroni}},
  \bibinfo{author}{\bibfnamefont{N.}~\bibnamefont{Bonini}},
  \bibinfo{author}{\bibfnamefont{M.}~\bibnamefont{Calandra}},
  \bibinfo{author}{\bibfnamefont{R.}~\bibnamefont{Car}},
  \bibinfo{author}{\bibfnamefont{C.}~\bibnamefont{Cavazzoni}},
  \bibinfo{author}{\bibfnamefont{D.}~\bibnamefont{Ceresoli}},
  \bibinfo{author}{\bibfnamefont{G.~L.} \bibnamefont{Chiarotti}},
  \bibinfo{author}{\bibfnamefont{M.}~\bibnamefont{Cococcioni}},
  \bibinfo{author}{\bibfnamefont{I.}~\bibnamefont{Dabo}}, \bibnamefont{et~al.},
  \bibinfo{journal}{J. Phys.: Condens. Matter} \textbf{\bibinfo{volume}{21}},
  \bibinfo{pages}{395502} (\bibinfo{year}{2009}).

\bibitem[{\citenamefont{Mizuguchi et~al.}(2009)\citenamefont{Mizuguchi,
  Tomioka, Tsuda, Yamaguchi, and Takano}}]{Mizuguchi2009JPSJ}
\bibinfo{author}{\bibfnamefont{Y.}~\bibnamefont{Mizuguchi}},
  \bibinfo{author}{\bibfnamefont{F.}~\bibnamefont{Tomioka}},
  \bibinfo{author}{\bibfnamefont{S.}~\bibnamefont{Tsuda}},
  \bibinfo{author}{\bibfnamefont{T.}~\bibnamefont{Yamaguchi}},
  \bibnamefont{and} \bibinfo{author}{\bibfnamefont{Y.}~\bibnamefont{Takano}},
  \bibinfo{journal}{J. Phys. Soc. Jpn.} \textbf{\bibinfo{volume}{78}},
  \bibinfo{pages}{074712} (\bibinfo{year}{2009}).

\bibitem[{\citenamefont{Xia et~al.}(2009{\natexlab{b}})\citenamefont{Xia, Hou,
  Zhao, Zhang, Chen, Luo, Wang, Wei, Lu, and Zhang}}]{Xia2009PRB}
\bibinfo{author}{\bibfnamefont{T.-L.} \bibnamefont{Xia}},
  \bibinfo{author}{\bibfnamefont{D.}~\bibnamefont{Hou}},
  \bibinfo{author}{\bibfnamefont{S.~C.} \bibnamefont{Zhao}},
  \bibinfo{author}{\bibfnamefont{A.~M.} \bibnamefont{Zhang}},
  \bibinfo{author}{\bibfnamefont{G.~F.} \bibnamefont{Chen}},
  \bibinfo{author}{\bibfnamefont{J.~L.} \bibnamefont{Luo}},
  \bibinfo{author}{\bibfnamefont{N.~L.} \bibnamefont{Wang}},
  \bibinfo{author}{\bibfnamefont{J.~H.} \bibnamefont{Wei}},
  \bibinfo{author}{\bibfnamefont{Z.-Y.} \bibnamefont{Lu}}, \bibnamefont{and}
  \bibinfo{author}{\bibfnamefont{Q.~M.} \bibnamefont{Zhang}},
  \bibinfo{journal}{Phys. Rev. B} \textbf{\bibinfo{volume}{79}},
  \bibinfo{pages}{140510} (\bibinfo{year}{2009}{\natexlab{b}}).

\bibitem[{\citenamefont{Kumar et~al.}(2010)\citenamefont{Kumar, Kumar, Saha,
  Muthu, Prakash, Patnaik, Waghmare, Ganguli, and Sood}}]{Kumar2010SSC}
\bibinfo{author}{\bibfnamefont{P.}~\bibnamefont{Kumar}},
  \bibinfo{author}{\bibfnamefont{A.}~\bibnamefont{Kumar}},
  \bibinfo{author}{\bibfnamefont{S.}~\bibnamefont{Saha}},
  \bibinfo{author}{\bibfnamefont{D.}~\bibnamefont{Muthu}},
  \bibinfo{author}{\bibfnamefont{J.}~\bibnamefont{Prakash}},
  \bibinfo{author}{\bibfnamefont{S.}~\bibnamefont{Patnaik}},
  \bibinfo{author}{\bibfnamefont{U.}~\bibnamefont{Waghmare}},
  \bibinfo{author}{\bibfnamefont{A.}~\bibnamefont{Ganguli}}, \bibnamefont{and}
  \bibinfo{author}{\bibfnamefont{A.}~\bibnamefont{Sood}},
  \bibinfo{journal}{Solid State Commun.} \textbf{\bibinfo{volume}{150}},
  \bibinfo{pages}{557 } (\bibinfo{year}{2010}), ISSN \bibinfo{issn}{0038-1098}.

\bibitem[{\citenamefont{Sugai et~al.}(2010)\citenamefont{Sugai, Mizuno, Kiho,
  Nakajima, Lee, Iyo, Eisaki, and Uchida}}]{Sugai2010PRB}
\bibinfo{author}{\bibfnamefont{S.}~\bibnamefont{Sugai}},
  \bibinfo{author}{\bibfnamefont{Y.}~\bibnamefont{Mizuno}},
  \bibinfo{author}{\bibfnamefont{K.}~\bibnamefont{Kiho}},
  \bibinfo{author}{\bibfnamefont{M.}~\bibnamefont{Nakajima}},
  \bibinfo{author}{\bibfnamefont{C.~H.} \bibnamefont{Lee}},
  \bibinfo{author}{\bibfnamefont{A.}~\bibnamefont{Iyo}},
  \bibinfo{author}{\bibfnamefont{H.}~\bibnamefont{Eisaki}}, \bibnamefont{and}
  \bibinfo{author}{\bibfnamefont{S.}~\bibnamefont{Uchida}},
  \bibinfo{journal}{Phys. Rev. B} \textbf{\bibinfo{volume}{82}},
  \bibinfo{pages}{140504} (\bibinfo{year}{2010}).

\bibitem[{\citenamefont{{Sugai} et~al.}(2010)\citenamefont{{Sugai}, {Mizuno},
  {Watanabe}, {Kawaguchi}, {Takenaka}, {Ikuta}, {Takayanagi}, {Hayamizu}, and
  {Sone}}}]{Sugai2010arXiv}
\bibinfo{author}{\bibfnamefont{S.}~\bibnamefont{{Sugai}}},
  \bibinfo{author}{\bibfnamefont{Y.}~\bibnamefont{{Mizuno}}},
  \bibinfo{author}{\bibfnamefont{R.}~\bibnamefont{{Watanabe}}},
  \bibinfo{author}{\bibfnamefont{T.}~\bibnamefont{{Kawaguchi}}},
  \bibinfo{author}{\bibfnamefont{K.}~\bibnamefont{{Takenaka}}},
  \bibinfo{author}{\bibfnamefont{H.}~\bibnamefont{{Ikuta}}},
  \bibinfo{author}{\bibfnamefont{Y.}~\bibnamefont{{Takayanagi}}},
  \bibinfo{author}{\bibfnamefont{N.}~\bibnamefont{{Hayamizu}}},
  \bibnamefont{and} \bibinfo{author}{\bibfnamefont{Y.}~\bibnamefont{{Sone}}}, 
  \eprint{arXiv:1010.6151v1} (unpublished).

\bibitem[{\citenamefont{Hu et~al.}(2008)\citenamefont{Hu, Dong, Li, Li, Zheng,
  Chen, Luo, and Wang}}]{Hu2008PRL}
\bibinfo{author}{\bibfnamefont{W.~Z.} \bibnamefont{Hu}},
  \bibinfo{author}{\bibfnamefont{J.}~\bibnamefont{Dong}},
  \bibinfo{author}{\bibfnamefont{G.}~\bibnamefont{Li}},
  \bibinfo{author}{\bibfnamefont{Z.}~\bibnamefont{Li}},
  \bibinfo{author}{\bibfnamefont{P.}~\bibnamefont{Zheng}},
  \bibinfo{author}{\bibfnamefont{G.~F.} \bibnamefont{Chen}},
  \bibinfo{author}{\bibfnamefont{J.~L.} \bibnamefont{Luo}}, \bibnamefont{and}
  \bibinfo{author}{\bibfnamefont{N.~L.} \bibnamefont{Wang}},
  \bibinfo{journal}{Phys. Rev. Lett.} \textbf{\bibinfo{volume}{101}},
  \bibinfo{pages}{257005} (\bibinfo{year}{2008}).

\bibitem[{\citenamefont{Chen et~al.}(2009)\citenamefont{Chen, Chen, Dong, Hu,
  Li, Zhang, Zheng, Luo, and Wang}}]{Chen2009PRB}
\bibinfo{author}{\bibfnamefont{G.~F.} \bibnamefont{Chen}},
  \bibinfo{author}{\bibfnamefont{Z.~G.} \bibnamefont{Chen}},
  \bibinfo{author}{\bibfnamefont{J.}~\bibnamefont{Dong}},
  \bibinfo{author}{\bibfnamefont{W.~Z.} \bibnamefont{Hu}},
  \bibinfo{author}{\bibfnamefont{G.}~\bibnamefont{Li}},
  \bibinfo{author}{\bibfnamefont{X.~D.} \bibnamefont{Zhang}},
  \bibinfo{author}{\bibfnamefont{P.}~\bibnamefont{Zheng}},
  \bibinfo{author}{\bibfnamefont{J.~L.} \bibnamefont{Luo}}, \bibnamefont{and}
  \bibinfo{author}{\bibfnamefont{N.~L.} \bibnamefont{Wang}},
  \bibinfo{journal}{Phys. Rev. B} \textbf{\bibinfo{volume}{79}},
  \bibinfo{pages}{140509} (\bibinfo{year}{2009}).

\bibitem[{\citenamefont{Han et~al.}(2009)\citenamefont{Han, Yin, Pickett, and
  Savrasov}}]{Han2009PRL}
\bibinfo{author}{\bibfnamefont{M.~J.} \bibnamefont{Han}},
  \bibinfo{author}{\bibfnamefont{Q.}~\bibnamefont{Yin}},
  \bibinfo{author}{\bibfnamefont{W.~E.} \bibnamefont{Pickett}},
  \bibnamefont{and} \bibinfo{author}{\bibfnamefont{S.~Y.}
  \bibnamefont{Savrasov}}, \bibinfo{journal}{Phys. Rev. Lett.}
  \textbf{\bibinfo{volume}{102}}, \bibinfo{pages}{107003}
  (\bibinfo{year}{2009}).

\bibitem[{\citenamefont{Han and Savrasov}(2009)}]{Han2009PRLa}
\bibinfo{author}{\bibfnamefont{M.~J.} \bibnamefont{Han}} \bibnamefont{and}
  \bibinfo{author}{\bibfnamefont{S.~Y.} \bibnamefont{Savrasov}},
  \bibinfo{journal}{Phys. Rev. Lett.} \textbf{\bibinfo{volume}{103}},
  \bibinfo{pages}{067001} (\bibinfo{year}{2009}).

\bibitem[{\citenamefont{Zhao et~al.}(2009)\citenamefont{Zhao, Adroja, Yao,
  Bewley, Li, Wang, Wu, Chen, Hu, and Dai}}]{Zhao2009NP}
\bibinfo{author}{\bibfnamefont{J.}~\bibnamefont{Zhao}},
  \bibinfo{author}{\bibfnamefont{D.~T.} \bibnamefont{Adroja}},
  \bibinfo{author}{\bibfnamefont{D.-X.} \bibnamefont{Yao}},
  \bibinfo{author}{\bibfnamefont{R.}~\bibnamefont{Bewley}},
  \bibinfo{author}{\bibfnamefont{S.}~\bibnamefont{Li}},
  \bibinfo{author}{\bibfnamefont{X.~F.} \bibnamefont{Wang}},
  \bibinfo{author}{\bibfnamefont{G.}~\bibnamefont{Wu}},
  \bibinfo{author}{\bibfnamefont{X.~H.} \bibnamefont{Chen}},
  \bibinfo{author}{\bibfnamefont{J.}~\bibnamefont{Hu}}, \bibnamefont{and}
  \bibinfo{author}{\bibfnamefont{P.}~\bibnamefont{Dai}},
  \bibinfo{journal}{Nature Phys.} \textbf{\bibinfo{volume}{5}},
  \bibinfo{pages}{555} (\bibinfo{year}{2009}).

\bibitem[{\citenamefont{Lumsden et~al.}(2010)\citenamefont{Lumsden,
  Christianson, Goremychkin, Nagler, Mook, Stone, Abernathy, Guidi, MacDougall,
  de~la Cruz et~al.}}]{Lumsden2010NP}
\bibinfo{author}{\bibfnamefont{M.~D.} \bibnamefont{Lumsden}},
  \bibinfo{author}{\bibfnamefont{A.~D.} \bibnamefont{Christianson}},
  \bibinfo{author}{\bibfnamefont{E.~A.} \bibnamefont{Goremychkin}},
  \bibinfo{author}{\bibfnamefont{S.~E.} \bibnamefont{Nagler}},
  \bibinfo{author}{\bibfnamefont{H.~A.} \bibnamefont{Mook}},
  \bibinfo{author}{\bibfnamefont{M.~B.} \bibnamefont{Stone}},
  \bibinfo{author}{\bibfnamefont{D.~L.} \bibnamefont{Abernathy}},
  \bibinfo{author}{\bibfnamefont{T.}~\bibnamefont{Guidi}},
  \bibinfo{author}{\bibfnamefont{G.~J.} \bibnamefont{MacDougall}},
  \bibinfo{author}{\bibfnamefont{C.}~\bibnamefont{de~la Cruz}},
  \bibnamefont{et~al.}, \bibinfo{journal}{Nature Phys.}
  \textbf{\bibinfo{volume}{6}}, \bibinfo{pages}{182} (\bibinfo{year}{2010}).

\bibitem[{\citenamefont{Sugai et~al.}(2003)\citenamefont{Sugai, Suzuki,
  Takayanagi, Hosokawa, and Hayamizu}}]{Sugai2003PRB}
\bibinfo{author}{\bibfnamefont{S.}~\bibnamefont{Sugai}},
  \bibinfo{author}{\bibfnamefont{H.}~\bibnamefont{Suzuki}},
  \bibinfo{author}{\bibfnamefont{Y.}~\bibnamefont{Takayanagi}},
  \bibinfo{author}{\bibfnamefont{T.}~\bibnamefont{Hosokawa}}, \bibnamefont{and}
  \bibinfo{author}{\bibfnamefont{N.}~\bibnamefont{Hayamizu}},
  \bibinfo{journal}{Phys. Rev. B} \textbf{\bibinfo{volume}{68}},
  \bibinfo{pages}{184504} (\bibinfo{year}{2003}).

\bibitem[{\citenamefont{Parkinson}(1969)}]{Parkinson1969JPC}
\bibinfo{author}{\bibfnamefont{J.~B.} \bibnamefont{Parkinson}},
  \bibinfo{journal}{Journal of Physics C: Solid State Physics}
  \textbf{\bibinfo{volume}{2}}, \bibinfo{pages}{2012} (\bibinfo{year}{1969}).

\bibitem[{\citenamefont{Chen et~al.}(2010)\citenamefont{Chen, Zhou, Zhang, Wei,
  Ou, Zhao, He, Ge, Arita, Shimada et~al.}}]{Chen2010PRB}
\bibinfo{author}{\bibfnamefont{F.}~\bibnamefont{Chen}},
  \bibinfo{author}{\bibfnamefont{B.}~\bibnamefont{Zhou}},
  \bibinfo{author}{\bibfnamefont{Y.}~\bibnamefont{Zhang}},
  \bibinfo{author}{\bibfnamefont{J.}~\bibnamefont{Wei}},
  \bibinfo{author}{\bibfnamefont{H.-W.} \bibnamefont{Ou}},
  \bibinfo{author}{\bibfnamefont{J.-F.} \bibnamefont{Zhao}},
  \bibinfo{author}{\bibfnamefont{C.}~\bibnamefont{He}},
  \bibinfo{author}{\bibfnamefont{Q.-Q.} \bibnamefont{Ge}},
  \bibinfo{author}{\bibfnamefont{M.}~\bibnamefont{Arita}},
  \bibinfo{author}{\bibfnamefont{K.}~\bibnamefont{Shimada}},
  \bibnamefont{et~al.}, \bibinfo{journal}{Phys. Rev. B}
  \textbf{\bibinfo{volume}{81}}, \bibinfo{pages}{014526}
  (\bibinfo{year}{2010}).

\bibitem[{\citenamefont{Devereaux and Hackl}(2007)}]{Devereaux2007RMP}
\bibinfo{author}{\bibfnamefont{T.~P.} \bibnamefont{Devereaux}}
  \bibnamefont{and} \bibinfo{author}{\bibfnamefont{R.}~\bibnamefont{Hackl}},
  \bibinfo{journal}{Rev. Mod. Phys.} \textbf{\bibinfo{volume}{79}},
  \bibinfo{pages}{175} (\bibinfo{year}{2007}).

\bibitem[{\citenamefont{Kato et~al.}(2009)\citenamefont{Kato, Mizuguchi,
  Nakamura, Machida, Sakata, and Takano}}]{Kato2009PRB}
\bibinfo{author}{\bibfnamefont{T.}~\bibnamefont{Kato}},
  \bibinfo{author}{\bibfnamefont{Y.}~\bibnamefont{Mizuguchi}},
  \bibinfo{author}{\bibfnamefont{H.}~\bibnamefont{Nakamura}},
  \bibinfo{author}{\bibfnamefont{T.}~\bibnamefont{Machida}},
  \bibinfo{author}{\bibfnamefont{H.}~\bibnamefont{Sakata}}, \bibnamefont{and}
  \bibinfo{author}{\bibfnamefont{Y.}~\bibnamefont{Takano}},
  \bibinfo{journal}{Phys. Rev. B} \textbf{\bibinfo{volume}{80}},
  \bibinfo{pages}{180507} (\bibinfo{year}{2009}).

\bibitem[{\citenamefont{Muschler et~al.}(2009)\citenamefont{Muschler, Prestel,
  Hackl, Devereaux, Analytis, Chu, and Fisher}}]{Muschler2009PRB}
\bibinfo{author}{\bibfnamefont{B.}~\bibnamefont{Muschler}},
  \bibinfo{author}{\bibfnamefont{W.}~\bibnamefont{Prestel}},
  \bibinfo{author}{\bibfnamefont{R.}~\bibnamefont{Hackl}},
  \bibinfo{author}{\bibfnamefont{T.~P.} \bibnamefont{Devereaux}},
  \bibinfo{author}{\bibfnamefont{J.~G.} \bibnamefont{Analytis}},
  \bibinfo{author}{\bibfnamefont{J.-H.} \bibnamefont{Chu}}, \bibnamefont{and}
  \bibinfo{author}{\bibfnamefont{I.~R.} \bibnamefont{Fisher}},
  \bibinfo{journal}{Phys. Rev. B} \textbf{\bibinfo{volume}{80}},
  \bibinfo{pages}{180510} (\bibinfo{year}{2009}).

\bibitem[{\citenamefont{Mazin et~al.}(2010)\citenamefont{Mazin, Devereaux,
  Analytis, Chu, Fisher, Muschler, and Hackl}}]{Mazin2010PRB}
\bibinfo{author}{\bibfnamefont{I.~I.} \bibnamefont{Mazin}},
  \bibinfo{author}{\bibfnamefont{T.~P.} \bibnamefont{Devereaux}},
  \bibinfo{author}{\bibfnamefont{J.~G.} \bibnamefont{Analytis}},
  \bibinfo{author}{\bibfnamefont{J.-H.} \bibnamefont{Chu}},
  \bibinfo{author}{\bibfnamefont{I.~R.} \bibnamefont{Fisher}},
  \bibinfo{author}{\bibfnamefont{B.}~\bibnamefont{Muschler}}, \bibnamefont{and}
  \bibinfo{author}{\bibfnamefont{R.}~\bibnamefont{Hackl}},
  \bibinfo{journal}{Phys. Rev. B} \textbf{\bibinfo{volume}{82}},
  \bibinfo{pages}{180502} (\bibinfo{year}{2010}).

\end{thebibliography}
\end{document}